\documentclass[fleqn,usenatbib,useAMS]{mnras}

\usepackage{graphicx,xcolor}	
\usepackage{amsmath}
\usepackage{amssymb}	
\usepackage{multicol}        
\usepackage{pdflscape}	
\usepackage[nameinlink,capitalise]{cleveref}

\usepackage[T1]{fontenc}
\usepackage{ae,aecompl}

\usepackage{newtxtext,newtxmath}
\usepackage{bm}		
\usepackage{ulem}

\def\eq#1{{Eq.~(\ref{#1})}}

\def\d {\mbox{d}}

\title[HI intensity power spectrum]{The HI intensity mapping power spectrum: \\ insights from recent measurements}

\author[Padmanabhan, Maartens, Umeh, Camera]{Hamsa Padmanabhan$^1$\thanks{hamsa.padmanabhan@unige.ch}, Roy Maartens$^{2,3,4}$, Obinna Umeh$^3$, Stefano Camera$^{5,6,7,2}$
\\~\\
$^{1}$D\'epartement de Physique Th\'eorique, Universit\'e de Gen\'eve 24, quai Ernest-Ansermet, CH
1211 Gen\'eve 4, Switzerland 
\\
$^2$Department of Physics \& Astronomy, University of the Western Cape, Cape Town 7535, South Africa\\
$^3$Institute of Cosmology \& Gravitation, University of Portsmouth, Portsmouth PO1 3FX, United Kingdom \\
$^4$National Institute for Theoretical \& Computational Sciences, Cape Town 7535, South Africa\\
$^5$Dipartimento di Fisica, Universit\'a degli Studi di Torino, Via P. Giuria 1, 10125 Torino,
Italy\\
$^6$Istituto Nazionale di Fisica Nucleare -- INFN, Sezione di Torino, Via P. Giuria 1, 10125
Torino, Italy\\
$^7$INAF -- Istituto Nazionale di Astrofisica, Osservatorio Astrofisico di Torino, 10025 Pino Torinese, Italy
}

\date{\today}

\pubyear{2024}

\begin{document}
\label{firstpage}
\pagerange{\pageref{firstpage}--\pageref{lastpage}}
\maketitle

\begin{abstract}
{The first direct measurements of the HI intensity mapping power spectrum were recently made using the MeerKAT telescope. These measurements are on nonlinear scales, at redshifts 0.32 and 0.44.
We develop a formalism for modelling small-scale power in redshift space, within the context of the mass-weighted HI halo model framework. 
This model is consistent with the latest findings from surveys on the HI-halo mass relation.
In order to model nonlinear scales,
we include the 1-halo, shot-noise and finger-of-god effects.
Then we apply the model to the MeerKAT auto-correlation data, 
finding that the model provides a good fit to the data at redshift 0.32, but the data may indicate some evidence for an adjustment at  $z \sim 0.44$. Such an adjustment can be achieved by an increase in the HI  {abundance or} halo model bias}.
\end{abstract}


\begingroup
\let\clearpage\relax
\endgroup

\section{Introduction}

The clustering of dark matter haloes and their tracers is usually described by its two-point correlation function and the associated power spectrum.
The small-scale structure of the power spectrum contains valuable information about baryonic effects and peculiar velocities which is useful for breaking degeneracies in cosmological parameters. Intensity mapping of neutral hydrogen (HI), a key tracer of cosmological dark matter, promises stringent constraints on the abundances and clustering of dark matter and baryons on cosmological scales.
So far, most studies focused on HI intensity mapping in cross-correlation with galaxy surveys using single-dish telescopes \citep[e.g.,][]{chang2010, masui13, switzer13, anderson2018, wolz2022} or telescope arrays operating in single-dish mode \citep[e.g.][]{cunnington2023}. The first direct detection and measurement of the HI intensity mapping auto-correlation fluctuations was recently made with the MeerKAT interferometer by \cite{paul2023}, {on nonlinear scales
$0.4< k \le 8\,{\rm Mpc}^{-1}$}.

In parallel, there has been recent progress on using stacked individual galaxy detections in HI to understand the properties of the baryon cycle out to $z \sim 0.3 - 1$, using the Giant Metrewave Radio Telescope \citep[GMRT;][]{bera2019, bera2022} and the MeerKAT MIGHTEE survey \citep{sinigaglia2022}. The results show that HI may play a significant role in the star formation process at these redshifts.

Previous work \citep{hparaa2017, hpar2017} has built a framework connecting the abundance and clustering of HI to that of dark matter in a data-driven manner, analogous to the halo model developed for the latter. In this framework, the small-scale structure of HI is modelled using the 1-halo term of the power spectrum, without considering the effect of peculiar velocities that alter the form of the power spectrum via redshift-space distortions (RSD). The RSD `finger-of-god'  effects are especially important on nonlinear scales. Theoretical prescriptions to model RSD in the context of HI have been applied to both hydrodynamical simulations \citep{villaescusa2018} and semi-analytic treatments populating HI in dark matter simulations \citep{sarkar2018}.

Preliminary findings from the cross-correlation of an HI intensity map from the Parkes telescope with a  galaxy sample from the 2dF survey at $z \sim 0$  \citep{anderson2018}, confirmed that RSD effects  become important at small scales.
 \citet{hprev2021} developed expressions for the small-scale structure of HI including redshift space effects, and found them to be a good match to the Parkes-2dF cross-correlation data. In particular, the low amplitude of clustering observed in the results was largely found to be {consistent with} an exponential finger-of-god suppression of power, within the HI halo-model framework.

Here, we expand upon the details of the derivation, formulating approximations for the RSD treatment in the halo-model framework for a generic, mass-weighted tracer.  We fix the parameters of the model to the best-fit values from stacking data, and  use a weighted average model for the finger-of-god damping parameter, following previous analyses in the dark matter context \citep{white2001, schaan2021}. { We also include a foreground wedge accounting for the non-spherical region of $k$-space in the measurement.}
Then we apply our model to the recent MeerKAT 
data on small scales. 
We find that at $z = 0.32$, the model provides a good fit to the data, including the small scale structure of the power spectrum. At the higher redshift interval centred at $z = 0.44$, the data may indicate evidence for a higher-than expected amplitude of the power spectrum compared to theoretical predictions. Such an increase may have different possible causes, including higher order biases or more complex baryonic effects. We also predict the two-dimensional power spectra at these redshifts from the formalism thus developed.

The paper is organised as follows. In \cref{sec:dmrsd}, we establish the notation and review the formalism for modelling the redshift-space power spectrum for dark matter in the halo model. We review the main ingredients involved in formulating the halo model for HI (and in general, a biased mass-weighted tracer of dark matter) in \cref{sec:hihalomodel}, including  the effects on small scales. We also compare the halo model predictions to recent results from the stacking of HI galaxies \citep{bera2022} and intensity mapping auto-correlation power spectra \citep{paul2023}.  In \cref{sec:hismallscales}, we formulate the treatment of RSD in the monopole of the HI power spectra analytically, and compare these to the MeerKAT results at $z = 0.32$ and $z= 0.44$,  { accounting for the non-spherical region of integration due to the foreground wedge in the measurement}.  We { describe the contributions of satellite galaxies to the small-scale HI power spectrum in \cref{sec:satps}, and} develop expressions for the two-dimensional power spectra in this framework in \cref{sec:2dps}.
We summarise our results and discuss the outlook in the concluding \cref{sec:discussion}.

\section{Fitting parameters for the {halo power spectrum in redshift space}}
\label{sec:dmrsd}
We need to model the small-scale behaviour of the HI intensity power spectrum, including RSD. 
In the halo-model formalism, the {halo} power spectrum is split into a 1-halo (correlations within the same dark matter halo) and a 2-halo (correlation between  perturbations in distinct halos) term \citep[see][for a review]{2002PhR...372....1C}:
\begin{equation}
    P_{\rm h} (k) = P_{\rm 1h}{(k)} + P_{\rm 2h}{(k)}\;,
\end{equation}
where
\begin{align}
P_{\rm 1h}{(k)} &=  \frac{1}{\bar{\rho}_{\rm m}^2} \int \d M \,  n(M)  M^2  |u_{\rm h} (k|M)|^2\;,
\label{1hdm}\\
P_{\rm 2h}{(k)} &=  P_{\rm lin} (k) \Big[\frac{1}{\bar{\rho}_{\rm m}} \int \d M \,  n(M)  M  b_{\rm h} (M)  \big|u_{\rm h} (k|M)\big| \Big]^2.
 \label{2hdm}
\end{align}
{Here $\bar{\rho}_{\rm m}$ is the average dark matter density, $M$ is the halo mass, $n$ is the halo mass function, $u_\mathrm{h}$ is the Fourier transform of the mass density profile,} $P_{\rm lin} (k)$ is the linear matter power spectrum, and $b_{\rm h}$ is the halo bias. In the above equations and elsewhere we have omitted the redshift dependence for brevity. { Throughout what follows, we assume the halo abundance $n(M)$ to be given by the Sheth and Tormen fitting form \citep{sheth2002} and the halo bias $b_{\rm h}$ to be given by that of \citet{scoccimarro2001}.}

Since observations make measurements of positions in redshift space, we modify the power spectra by following, e.g.\ the treatments in \citet{seljak2000}, \citet{white2001}, and \citet{peacock1994}. The position of a  halo in redshift space is
\begin{equation}
    \bm{s} = \bm{x} + \frac{\bm{v}\cdot \hat{\bm n} }{a\,H}\,\hat{\bm n}\;,
\end{equation}
where $\bm{x}$ is the position in real space,  $\bm{v}\cdot \hat{\bm n}$ is the velocity component along the line-of-sight direction $\hat{\bm n}$ (assumed to be fixed), and $H$ is the Hubble rate at scale factor $a$. As a result, the halo density field in Fourier space, $\delta_{\rm h}$, acquires a correction { which can be phenomenologically described in the `dispersion model' \citep{scoccimarro2004} by the form:}
\begin{equation}
    \delta_{\rm h}^{\rm s}(\bm k) = \delta_{\rm h}(\bm k)\Big(1 + f\mu^2\Big) \exp\Big(-\frac{1}{2} k^2\mu^2\sigma^2\Big) ,
\end{equation}
made up of a Kaiser \citep{kaiser1987} and a finger-of-god term,\footnote{ The model above is technically a simplification since the dispersion term and the large-scale velocity field are not independent, and the dispersion itself is typically a function of scale and halo bias. Nevertheless, it has been found to describe the redshift space power spectrum of dark matter out to quasilinear scales, including the suppression seen in $N$-body simulations \citep{jing2001, white2001}.}
where 
\begin{equation}
k=|\bm k|= \Big({k_\parallel^2+\bm k_\perp^2}\Big)^{1/2}\,,~~\mu = \hat{\bm k}\cdot \hat{\bm n}=\frac{k_{\parallel}}{k}\,, ~~ f=\frac{\d \ln D}{\d \ln a}\,.   
\end{equation}
Here $D$ is the linear growth factor whose growth rate can be well approximated as $f(a) \simeq {[\Omega_\mathrm{m}(a)]^{0.55}}$.
The finger-of-god damping is described by a Gaussian (exponential) suppression, where $\sigma$ is the pairwise velocity dispersion (which is a function of halo mass and redshift).\footnote{An alternative  model is the Lorentzian function
$(1+k^2 \mu^2 \sigma^2/2)^{-1}$.}

Then the RSD power spectrum is
    $P_{\rm h}^{\rm s} = P^{\rm s}_{\rm 1h}+ P^{\rm s}_{\rm 2h}$, {where} 
\begin{align}
P^{\rm s}_{\rm 1h}(k,\mu) &=   \frac{\big(1+f\mu^2\big)^2}{\bar{\rho}_{\rm h}^2}\!\int \!\d M   n(M)  M^2 
%\notag \\&\times 
\big|u_{\rm h} (k|M)\big|^2  {\rm e}^{-k^2\mu^2\sigma^2},
\label{1hrsddm}\\
P^{\rm s}_{\rm 2h}(k,\mu) &=  P_{\rm lin}(k) \bigg[\frac{\big(1+f\mu^2\big)}{\bar{\rho}_{\rm h}}
\notag\\
&~~~ \times \int\! \d M   n(M)  M  b_{\rm h} (M) 
%\ \nonumber \right. \\&\times \left. 
\big|u_{\rm h} (k|M)\big|\,{\rm e}^{-k^2\mu^2\sigma^2/2} \bigg]^2\!.
\label{2hrsddm}
\end{align}
The monopole is the average power spectrum ${\bar{P}}_{\rm h}^{\rm s} =\int \d\mu\, P_{\rm h}^{\rm s}/2$, which leads to:
\begin{align}
    \bar {P}^{\rm s}_{\rm 1h} (k) &=   \frac{1}{\bar{\rho}_{\rm h}^2} \int \d M \,  n(M)  M^2  \big|u_{\rm h} (k|M)\big|^2 \ \mathcal{R}_1 (k\sigma),\\
\bar{P}^{\rm s}_{\rm 2h}(k) &\simeq \left( 1+\frac23 f + \frac15 f^2\right)P_{\rm lin}(k)\notag \\
&~~~\times \left[\frac1{\bar{\rho}_{\rm h}} \int \d M \,  n(M)  M  b_{\rm h} (M)  \mathcal{R}_2(k\sigma)  \big|u_{\rm h} (k|M)\big| \right]^2.
\label{pdmrsd1h}
\end{align}
The functions ${\cal R}_i$ are given by\footnote{The approximation made in \cref{pdmrsd1h} is that the $\mu$-integral of the Gaussian term is performed within the $M$-integral, even though it should technically be done along with the one for the Kaiser term. As pointed out by \citet{white2001}, this is nevertheless  found to be in good agreement with simulations \citep{seljak2000}.}
\begin{align}
   {\cal R}_1(y) &= \frac12\int_{-1}^1 \d\mu \left(1+f\mu^2\right)^2 {\rm e}^{-{y}^2\mu^2}   = \frac{\sqrt{\pi}}{8}\,\frac{ {\rm erf}(y)}{y^5}\,
    \Big[ 3f^2+
    \nonumber \\
   &~~~~~
    4fy^2+4y^4 \Big]  - \frac{{\rm e}^{-y^2}}{4y^4}\Big[ f^2\big(3+2y^2\big)+4fy^2 \Big] ,\nonumber \\
   {\cal R}_2(y) &= \frac12\int_{-1}^1 \d\mu \,{\rm e}^{-{y}^2\mu^2/2}   %\nonumber \\ &
   = \sqrt{\frac\pi2}\,\frac{ {\rm erf}(y/\sqrt2\,)}y.
   \label{rispherical}
\end{align}
Note that the $M$-dependence in ${\cal R}_i$ comes from the $\sigma$ term.\footnote{Our definitions of ${\cal R}_1(y)$ and ${\cal R}_2(y)$ are reversed compared to those in \citet{white2001}.}
Similar expressions hold for the Lorentzian finger-of-god damping factor.

\section{Halo model for HI and its parametrisation}
\label{sec:hihalomodel}

\begin{figure}
    \centering
    \includegraphics[width = \columnwidth]{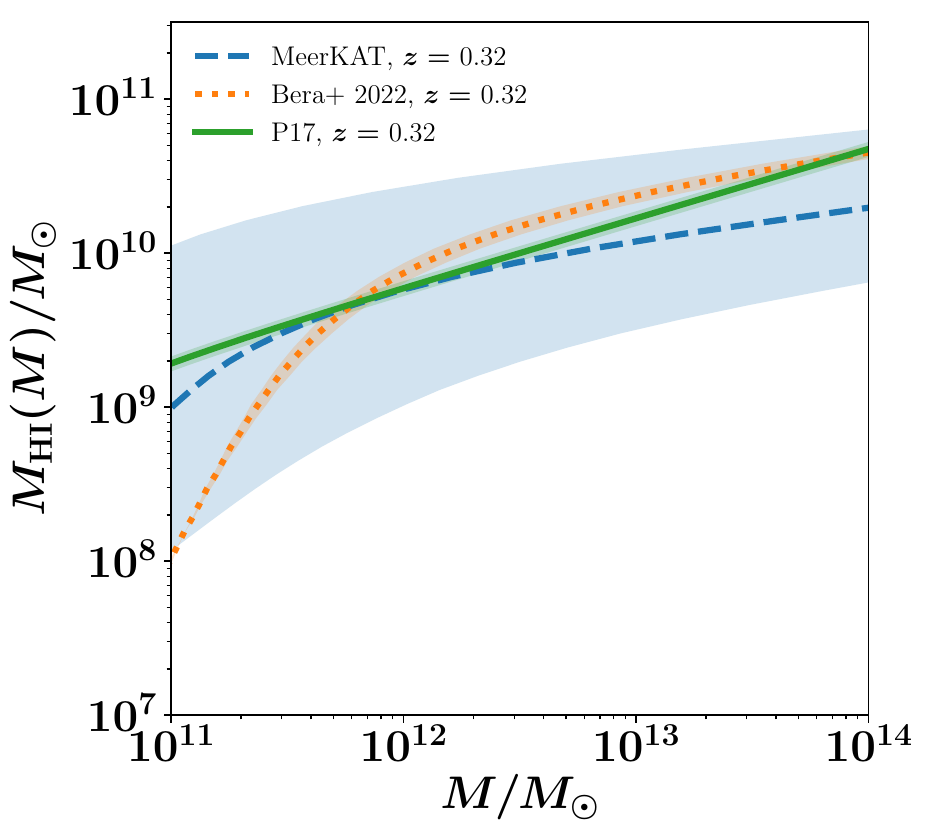}
    \caption{Halo model predictions for the average HI mass - halo mass relation at $z = 0.32$ {(green solid line)}, compared to the relation inferred from abundance matching the findings of the MeerKAT intensity mapping observations \citep{paul2023} (blue dashed line) and the galaxy stacking results of \citet{bera2022} (orange dotted line). { The shaded blue and orange regions denote the uncertainties in the relation inferred from those in the parameters of the luminosity function of the surveys \citep{paul2023, bera2022} and in the HI-halo mass relation \citep{hparaa2017}.}
    }
    \label{fig:hihm}
\end{figure}

We now extend this formalism to describe the small-scale structure of the power spectrum of a mass-weighted tracer. In the case of HI, we start with the formalism connecting { the average HI mass in a dark matter halo of mass $M$ at redshift $z$, denoted by $M_{\rm HI}(M,z)$ and described by \citep{hparaa2017}:
\begin{equation}
M_{\rm HI} (M,z) = \alpha f_{\rm H,c} M \left(\frac M{10^{11}\,h^{-1}{\rm M}_\odot}\right)^{\beta} \!
\exp\left\{\!-\left[\frac{v_{{\rm c},0}}{v_{\rm c}(M,z)}\right]^3\right\},
\label{hihm}
\end{equation}
In the above, the circular velocity $v_{\rm c} (M,z)$ is defined in terms of the virial radius $R_{\rm v}(M,z)$ of the halo, 
 \begin{equation}
    v_{c}(M,z) = \sqrt\frac{GM}{R_{\rm v}(M,z)}
    \label{vcRv}
\end{equation}
with
\begin{equation}
 R_{\rm v} (M,z) = 46.1 \ {\rm{kpc}} \  \left(\frac{\Delta_v \Omega_m h^2}{24.4} \right)^{-1/3} \left(\frac{1+z}{3.3} \right)^{-1} \left(\frac{M}{10^{11} M_{\odot}} \right)^{1/3}
 \label{virialradius}
 \end{equation} 
The cosmological hydrogen fraction is denoted by $f_{\rm H,c}$ and given by
\begin{equation}
f_{\rm H,c} = (1 - Y_{\rm He}) \Omega_b/\Omega_m \, ,
\label{fhc}
\end{equation}
where $Y_{\rm He} = 0.24$ denotes the primordial helium fraction.}
The three free parameters $\alpha$, $\beta$ and $v_{\rm c,0}$ describe the overall normalization { of HI in the halo,  relative to the cosmic fraction of hydrogen}, the logarithmic slope and lower virial velocity cutoff of the dark matter halo hosting HI, respectively.  Their best-fitting values and $1\sigma$ uncertainties, constrained by a joint fit to the available HI data including the statistics of Damped Lyman Alpha systems (DLAs) and HI galaxy surveys \citep{hpar2017, hparaa2017}, were found to be given by:
\begin{equation}
\alpha = 0.09 \pm 0.01,~ 
\beta = -0.58 \pm 0.06,~
\log_{10} \Big[\frac{v_{{\rm c},0}}{1\,\mathrm{km/s}}\Big] = 1.56 \pm 0.04.
\label{hihmbestfit}
\end{equation}
We compare the HI mass -- halo mass relation predicted by the halo model to recent observations in \cref{fig:hihm}. We consider two recent surveys targeting HI in galaxy surveys and intensity mapping at $z \sim 0.3-0.4$: (1)~the MeerKAT survey \citep{paul2023}, in which HI is probed down to very small scales, $k \sim 10\,h\,\mathrm{Mpc}^{-1}$; and (2)~recent results probing stacked 21 cm emission in star-forming galaxies at $z \sim 0.35$ from the Giant Metrewave Radio Telescope (GMRT) Extended Goth Strip (EGS) survey \citep{bera2019, bera2022}. These and recent results probing HI in the $z \sim 1$ range  \citep{chowdhury2021, chowdhury2022} suggest that HI may evolve strongly over $z \sim 0-1$ and even contribute a major fraction of fuel for the star-formation.

The mass functions from both surveys, denoted by $\phi(M_{\rm HI})$, are used to constrain the HI mass -halo mass relation via the abundance matching technique \citep[e.g.][]{hpgk2017}, which assumes a monotonicity between the HI mass and the underlying halo mass:
\begin{equation}
  \int_{M(M_{\rm HI})}^{\infty}\d \log_{10} \! M\, \frac{\d n}{ \d \log_{10}\! M}  = \int_{M_{\rm HI}}^{\infty}\d \log_{10}\! M_{\rm HI}\, \phi(M_{\rm HI}) \;.
  \label{abmatchcii}
\end{equation}
{ The shaded areas on the blue, orange and green curves denote, respectively, the range of halo masses admissible as a function of the HI mass, resulting from the uncertainties in the $\phi(M_{\rm HI})$ inferred from the  MeerKAT intensity mapping  survey \citep{paul2023} and the GMRT stacking analysis \citep{bera2022}, as well as the model parameters in the HI to halo mass relation \citep{hparaa2017}. }
\Cref{fig:hihm} shows that the current halo model parameters {(green curve)} are consistent with the latest MeerKAT findings {(blue curve)}, while the stacking results {(orange curve)} show a steeper drop at low HI masses. This is in line with theoretical expectations, since the stacking results target individual galaxies while the MeerKAT result additionally measures the overall or integrated emission from lower luminosity sources. 

The small-scale structure of HI is encoded in the {density} profile, given by
\begin{equation}
\rho_{\rm HI} (r,M,z) = \rho_0 \,\exp\left[-\frac{r}{r_{\rm s}(M,z)}\right]\;,\label{rhodefexp}
\end{equation}
in which $r_{\rm s}$ is the scale radius of the halo, and defined as $r_{\rm s}(M,z)\equiv R_{\rm v}(M,z)/c_{\rm HI}(M,z)$. Here, $R_{\rm v}(M, z)$ denotes the virial radius of the halo given by \cref{virialradius}, and $c_{\rm HI}$ the concentration of the HI, analogous to the corresponding expression for dark matter, defined as \citep{maccio2007}
\begin{equation}
c_{\rm HI}(M,z) = c_{{\rm HI},0} \left(\frac{M}{10^{11}\, {\rm M}_\odot} \right)^{-0.109} \frac4{(1+z)^\gamma}\;.
\end{equation}
The constant of proportionality in \cref{rhodefexp}, denoted by $\rho_0$, is found by normalising the HI mass within the virial radius $R_{\rm v}(M,z)$ at a given halo mass and redshift to $M_{\rm HI}$. This is found to be well-approximated by:
\begin{equation}
\rho_0(M,z) = \frac{M_{\rm HI} (M)}{8\pi r_{\rm s}(M,z)^3}\;.\label{eq:rho0}
\end{equation}
{ Strictly speaking, the profile is truncated at the virial radius of the host halo, $R_{\rm v}(M,z)$ rather than at  infinity, as required by the normalization above. However, this difference is negligible in practice, due to the steep fall-off of the function \eq{rhodefexp} with $r$.}
The radial density profile thus introduces two more free parameters: the normalization of the concentration, $c_{\rm HI,0}$, and its redshift-dependent slope, $\gamma$. The best-fitting values of these parameters were found to be:
\begin{equation}
c_{\rm HI, 0} = 28.65 \pm 1.76,~ 
\gamma = 1.45 \pm 0.04
\label{rhobestfit}
\end{equation}
in the HI halo model framework \citep{hparaa2017}.
    
Using this parametrised form for the HI-halo mass relation, we compute the power spectrum on both linear and nonlinear scales from
\begin{align}
P_{\rm 1h, HI}{(k)} &=  \frac{1}{\bar{\rho}_{\rm HI}^2}\, \int \d M \,  n(M)  M_{\rm HI}^2  \big|u_{\rm HI} (k|M)\big|^2,
\\
P_{\rm 2h, HI}{(k)} &=  P_{\rm lin} (k) \!\left[\frac{1}{\bar{\rho}_{\rm HI}}\! \int \!\d M   n(M)  M_{\rm HI} (M)  b_{\rm h} (M) \big|u_{\rm HI} (k|M)\big| \right]^2\!\!,
\end{align}
where 
the terms are defined analogously to \cref{1hdm,2hdm}.
The mean HI density is given by
\begin{equation}
\bar{\rho}_{\rm HI} = \int \d M\, n(M) M_{\rm HI} (M) \;,
\end{equation}
and the normalised Fourier transform of the HI density profile is given by
\begin{equation}
u_{\rm HI}(k|M) = \frac{4 \pi}{M_{\rm HI} (M)} \int_0^{R_{\rm v} (M)}\d r\,r^2 \rho_{\rm HI}(r) \,\frac{\sin kr}{kr} \;.
\label{uhi}
\end{equation}

{ Some comments about the nature and  ingredients of the framework  above are in order, which will also be useful in the extensions described in the next Sections. As can be seen from Eqs. (\ref{hihm}), (\ref{rhodefexp}) and (\ref{uhi}), the approach describes a continuous distribution of HI in haloes, analogously to the corresponding formalism for dark matter. 

As such, the framework at hand lends itself naturally to a description of intensity mapping, where the targets are unresolved HI intensities rather than resolved galaxies. This differs from a Halo Occupation Distribution (or HOD) description in which HI is assumed to reside in discrete central or satellite galaxies (and where one is explicitly interested in the number counts of each). The present approach also has the advantage of being able to capture that part of 21 cm  emission which need not be correlated with real-space positions of galaxies \citep[similar approaches have been adopted to describe the diffuse infrared emission from galaxies and the intra-cluster medium, for a recent example see][]{maniyar2021}.

In the context of the mass-weighted tracer-halo model framework, the technique of abundance matching is used to determine the average mass  of the tracer (in this case, HI) in a given mass bin of dark matter [see, e.g., \citet{hpgk2017, hpco, hpcii,  chung2022,hp2023} for examples of applying this technique to continuous distributions of HI mass and tracer line luminosity in the context of intensity mapping.] Again, this differs from the abundance matching of galaxy luminosity functions, where both the central and satellite number counts need to be matched separately \citep[see, e.g.,][]{vale2004}. { The HI to halo mass relation  in its current form (Eqs. \ref{hihm} - \ref{hihmbestfit}) is well fitted to the stacking results of HI galaxy surveys and the DLA incidence at $z \sim 0.3-0.6$ \citep{hptrcar2016, hpar2017} in the redshift range of interest.}

 The form of the profile favoured by the observations (Eqs. \ref{rhodefexp} - \ref{rhobestfit}) is highly concentrated, translating loosely into a smaller contribution to the bulk distribution of HI from satellites as compared to central galaxies. The parameters describing the small-scale behaviour of the profile function are constrained by matching to several independent datasets covering the redshift range of interest, including the column density distribution of HI from the WHISP survey \citep{zwaan2005a} and the clustering of
HI galaxies from the ALFALFA survey  \citep{martin12}, both at $z \sim 0$, the DLA column density and incidence of DLAs at $z \sim 0.6$ and $z \sim 1.2$ \citep{rao06} and at higher redshifts, $z \sim 2.3$ \citep{noterdaeme12}. The profile is especially found to be a good fit to the small-scale clustering at $z \sim 0$ [Fig. 2 of \citet{hparaa2017}], a result that carries over to the mass-weighted case as the scale dependence of clustering strength of HI galaxies is found to be fairly independent of halo mass (Matteo Nicoli, private communication). }

Another contribution to the HI power spectrum on small scales comes from the shot noise \citep{villaescusa2018}, { which, in the intensity mapping context is given by the second moment of the observed field \citep[e.g.,][]{breysse2017, kovetz2017}:}
\begin{equation}
   P_{\rm SN} =  \frac{1}{\bar{\rho}_{\rm HI}^2} \int \d M \,  n(M)  M_{\rm HI}^2 \;. 
   \label{shotnoiseHI}
\end{equation}
This term corresponds to the $k \to 0$ limit of the 1-halo power spectrum, and provides a measure of the  Poisson noise coming from the finite number of discrete sources \citep{seo2010, seehars2016}. 
{ The shot noise term is, in some cases, added to the sum of the 1- and 2-halo contributions, even though it is technically double-counting if interpreted as the $k \to 0$ limit of the 1-halo term. In a number-counts weighted interpretation, the two are equivalent in the case when the satellite contributions are negligible \citep{ginzburg2017}. Physically, the shot noise and 1-halo terms capture two different aspects of the astrophysics in line-intensity mapping: with the 1-halo term being sensitive to the mean value of tracer luminosity (or mass) squared and the shot noise being a measure of  its mean squared value \citep[e.g.,][]{schaan2021a}. While a continuous tracer-to-halo  framework does not technically accommodate a Poisson-noise like contribution, it is worth noting that  the profile function in \eq{rhodefexp} implicitly encodes the contribution of `satellites' around the galaxy weighted by their HI mass.  

As far as the power spectrum is concerned, it has been shown using simulations \citep[e.g.][]{wolz2019} that the auto-power spectra predicted by the discrete (i.e. number-counts weighted) and continuous (mass-weighted) distributions differ only on scales $k > 10 h$ Mpc$^{-1}$, indicating that the contributions of the central-satellite term (and thus that of the satellites themselves) enter only on much smaller scales beyond the regime probed by current experiments. Furthermore, in the halo mass regime of $ < 10^{12} M_{\odot}$,  the mean number of satellite galaxies is expected to be {negligibly small \citep[][see Figs. 1-4]{zhang2022}, less than a few percent of the
the mean number of centrals over $z \sim 0 - 1$. This implies a negligible contribution of the satellites to the HI power spectrum on scales 1 h Mpc$^{-1} < k < 10$ h Mpc $^{-1}$, as also seen by the power spectrum obtained by \citet{baugh2019} including the  contribution of satellite galaxies (Fig. 10 of that paper) compared to our present results on these scales. We use the recent ALFALFA results at $z \sim 0$ to explicitly estimate the magnitude of the contribution of satellites to the small-scale HI power spectrum in Appendix \ref{sec:satps}, reaching similar conclusions.}

\section{Contributions to the power spectrum at small scales}
\label{sec:hismallscales}

The power spectrum, both in one- and two-dimensional $k$-space, contains information on the HI distribution on small scales. This is made up of the contributions from (1)~the shot noise,  (2)~the 1-halo term described in the previous section, and (3)~the RSD finger-of-god  effects. 

We now consider the contributions of RSD affecting the HI distribution on small scales. To do this, we generalise the expressions found in \cref{sec:dmrsd} to the case of HI in the context of the halo model for a mass-weighted, biased tracer.  In the presence of RSD, the density of HI (in Fourier space) is boosted on large scales by the Kaiser term:
\begin{equation}
    \delta_{{\rm HI}}^{\rm s}({\bm k}) = \delta_{{\rm HI}}({\bm k}) +  f \mu^2 \delta_{{\rm h}}({\bm k})\,,
\end{equation}
where $f \delta_{{\rm h}}$ denotes the velocity divergence of the dark matter. {  In the above, the Kaiser effect
is applied to the underlying halo density field since it is sourced by the large-scale velocities associated with the halo}. In order to incorporate the finger-of-god effect on small scales,\footnote{The present prescription employed for the finger-of-god effects was found to be a good fit to the Parkes HI cross-correlation data \citep{hprev2021}.} we damp the previous expression: 
\begin{equation}
   \delta_{{\rm HI}}^{\rm s} \to \delta_{{\rm HI}}^{\rm s} \exp \left(-\frac{1}{2} k^2 \mu^2 \sigma^2\right)\,.
\end{equation}
Using the three equations above gives the full expression for the anisotropic ($\mu$-dependent) power spectrum of HI:
\begin{align}
   P_{\rm HI}^{\rm s}(k, \mu) =&\;  \left\langle \left| \delta_{{\rm HI}}(\bm k)  \exp\left(-\frac{1}{2} k^2  \mu^2  \sigma^2\right)\right|^2 \right\rangle \nonumber \\
   &+  \left\langle \left| f  \mu^2  \delta_{{\rm h}}(\bm k) \, \exp\left(-\frac{1}{2} k^2  \mu^2  \sigma^2\right)\right|^2 \right\rangle \nonumber \\
   &+ 2  \left\langle f  \mu^2  \delta_{{\rm {h}}}(\bm k)  \delta_{{\rm HI}}(\bm k)  \exp\left(-k^2  \mu^2  \sigma^2\right) \right\rangle\,.
   \label{phirsd}
\end{align}
Applying the same procedure as for the dark matter only case treated in \cref{sec:dmrsd}, 
the monopole of the HI power spectrum in redshift space can be expressed by
\begin{align}
\bar{P}_{\rm HI}^{\rm s}(k) &= \left( F_{\rm HI}^2+\frac23 F_{\rm m}F_{\rm HI} + \frac15 F_{\rm m}^2\right) P_{\rm lin}(k) \notag\\
&+ \frac1{\bar{\rho}_{\rm HI}^2}
\int \d M \, n(M)  M_{\rm HI}^2 \,
{\cal R}_1(k\sigma)  \big|u_{\rm HI}(k,M)\big|^2,
\label{rs}
\end{align}
where 
\begin{align}
F_{\rm m}&=f\int \d M \, n(M)  b(M) \, {\cal R}_2(k\sigma)  u_{\rm h}(k, M), \nonumber\\
F_{\rm HI}&=
\frac1{\bar{\rho}_{\rm HI}}  \int \d M \,
n(M)  M_{\rm HI}(M)  {b_{\rm h}}(M)\,  {\cal R}_2(k\sigma)  u_{\rm HI} (k,M)\;.
\label{p0}
\end{align}
The above expressions generalise the treatment for dark matter to the case of its mass-weighted, biased tracers. Note that there is no further correction to the shot noise term \cref{shotnoiseHI} in this approach as it is taken to be the $k \to 0$ limit of the 1-halo term.

{In the above expressions, the line-of-sight velocity dispersion is given by $\sigma(M,z) \equiv \sigma_{v, \rm 1D}(M, z)(1+z)/H(z)$, where $\sigma_{v, \rm 1D}(M, z)$  follows that of a singular isothermal sphere of mass $M$ and radius $R_{\rm v}(M, z)$,  e.g. \citet{white2001}:
\begin{equation}
    \sigma_{v, \rm 1D}(M, z)^2 =  \frac{GM}{2 R_{\rm v}(M, z)}\;,
\end{equation}
This form is found to be a good match to the results of simulations \citep{evrard2008}.}
The resultant dispersion of the effective displacement in redshift space due to
the random line-of-sight motion within halos is denoted by $\sigma_{\rm avg}$.
This quantity is modelled as the mass-weighted average of its per-halo value, weighted by a factor of $1/ a H$  \citep[e.g.,][]{schaan2021}:
\begin{equation}\label{savg}
    \sigma_{\rm avg}(z) = \frac{\int \d M\, M n (M,z) \sigma_{v, \rm 1D} (M,z)}{\int  \d M\, M n (M,z)}\,   \frac{(1+z)}{{H}(z)}\; \, ,
\end{equation}
in a manner similar to the effective HI bias \citep{hpar2017}. {Note that the $ \sigma_{\rm avg}(z) $ as defined above is expressed in units of length. Numerically, we find $\sigma_{\rm avg} \sim 225\ {\rm km/s} \  (1+z)/H(z) \sim 3.5$ Mpc at $z \sim 0.3 - 0.4$.}

The monopole of the power spectrum from \cref{rs} including the RSD is plotted as the solid thick lines in \cref{fig:psrsd} as a function of redshift and scale. The dashed thin lines show the power spectra without the RSD effects.
\begin{figure}
    \centering
    \includegraphics[width = \columnwidth]{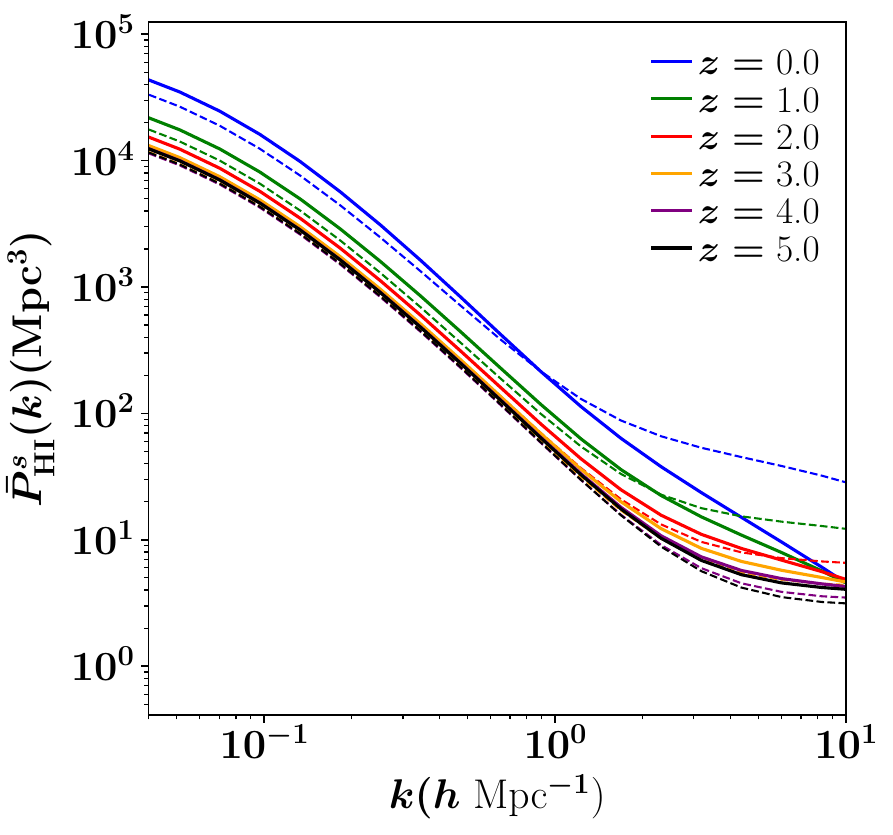}
    \caption{Power spectrum {monopole} of HI including (solid thick lines) and without (dashed thin lines) the effect of RSD, modelled by \cref{rs}.
    }
    \label{fig:psrsd}
\end{figure}

\begin{figure}
    \centering
    \includegraphics[width = \columnwidth]{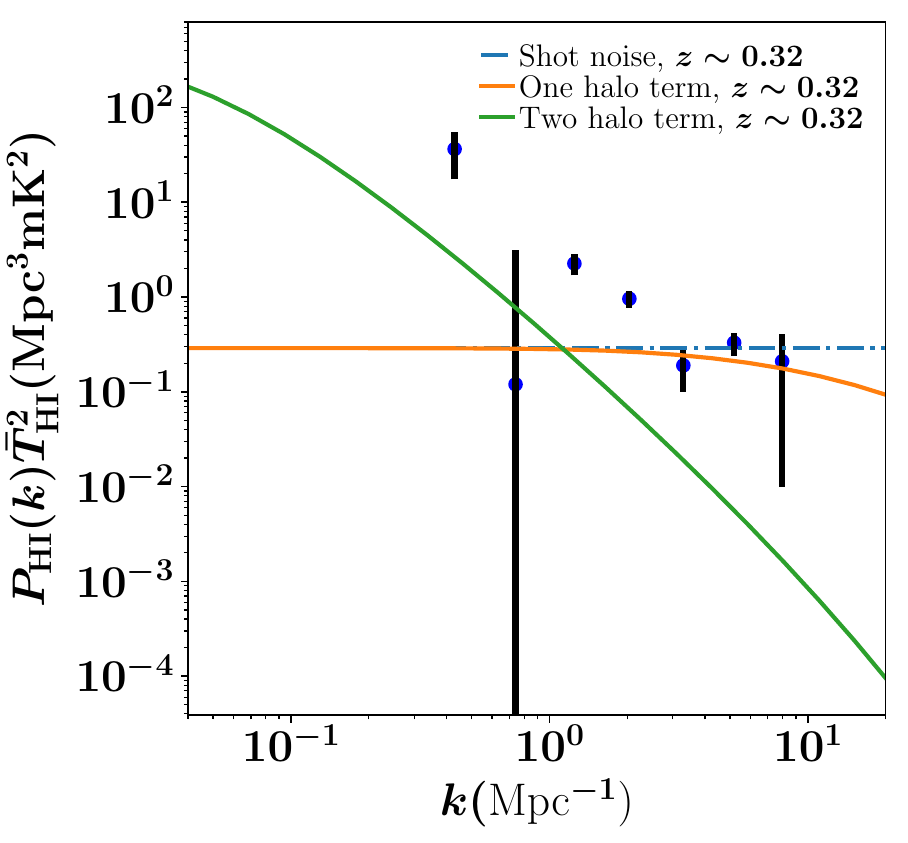}
    \caption{ Contributions to the  HI power spectrum at $z \sim 0.32$ at large and small scales: the 1- and 2-halo terms and the shot noise { described by \eq{shotnoiseHI}}. The points with error bars show the measurements from the MeerKAT intensity mapping survey at the same redshift.}
    \label{fig:hicontribs}
\end{figure}

\begin{figure}
    \centering
    \includegraphics[width = \columnwidth]{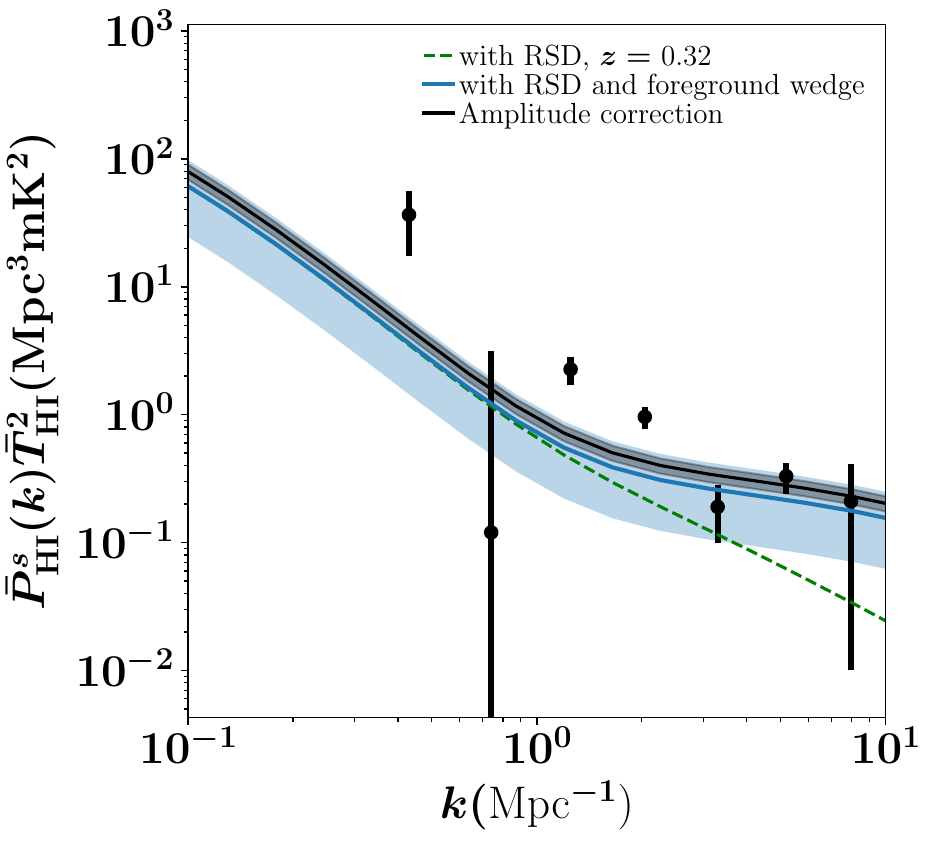}
    \includegraphics[width = \columnwidth]{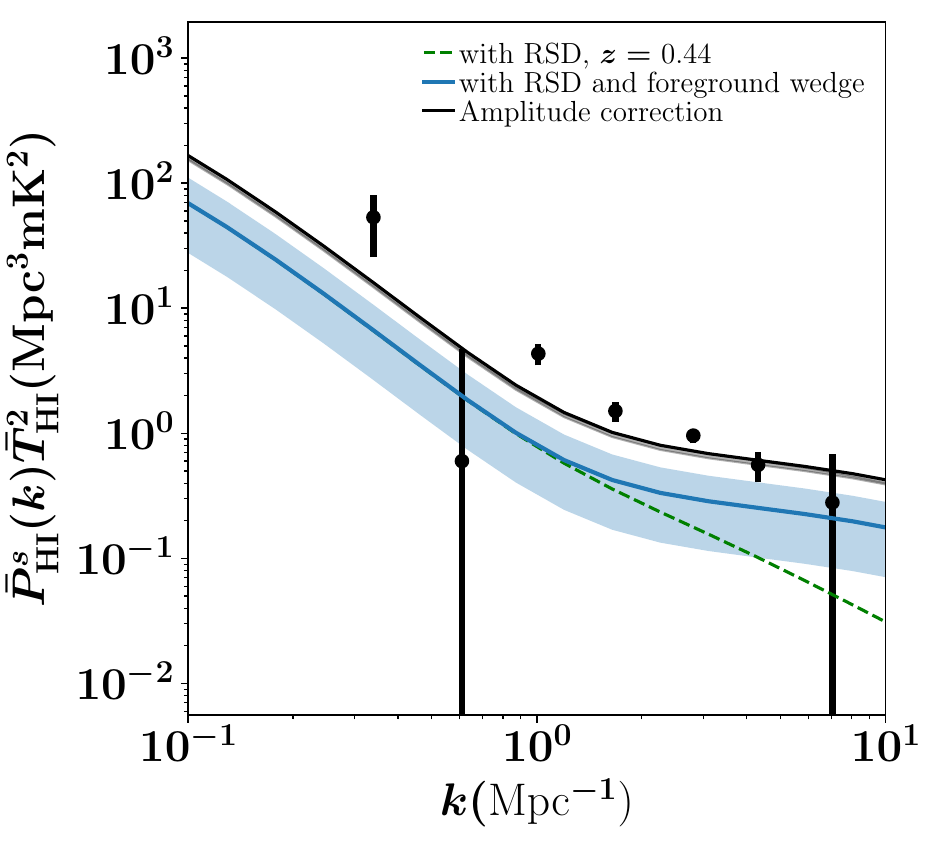}
    \caption{ The power spectra  $\bar{P}_{\rm HI}^{\rm s}(k)$ given by \eq{rs}, normalized to temperature units by \eq{tempnormHI} at $z = 0.32$ (top panel) and $z = 0.44$ (lower panel) are shown by the green dashed lines. The power spectra including the foreground wedge with $A = 0.3$ and the one-halo term unmodulated by finger-of-god effects [\eq{rsfw}] are shown by the blue curves. The expected scatter due to astrophysical uncertainties \citep{hprev2021} is shown by the blue shaded band.
 The MeerKAT intensity mapping results are shown by the black points with error bars. The best-fitting curve to the data, leaving free a scaling to the amplitude of the power spectrum, is shown by the black solid line, with the associated uncertainty in the inferred amplitude parameter indicated by the grey band.}
    \label{fig:psrsdmeerkat}
\end{figure}

We now illustrate the comparison of these findings to the recent MeerKAT intensity mapping observations.  { Since the observations are in temperature units,  we scale the power spectra found in the previous section with the mean HI temperature, defined by \citep[e.g.,][]{battye2012, camera2013, bull2014}:
\begin{eqnarray}
\bar{T}_{\rm HI} &=& \frac{3 h c^3 A_{10}}{32 \pi \nu_{21}^2 k_B} \frac{(1+z)^2}{H(z)} n_{\rm HI} \nonumber \\
&=& 566 h \frac{H_0}{H(z)} \frac{\Omega_{\rm HI}(z)}{0.003} (1+z)^2 \ \mu {\rm  K}
\label{tempnormHI}
\end{eqnarray}
where $n_{\rm HI}$ is the comoving number density of HI, and
\begin{equation}
\Omega_{\rm HI}(z) = \frac{\rho_{\rm HI} (1+z)^{-3}}{\rho_{c,0}}
\end{equation}
in which $\rho_{\rm HI}$ is the proper (mass) density of HI. Most observations [to which the halo model described in the previous section is fitted, as well as the recent results in \citet{bera2022}] favour  $\Omega_{\rm HI} = 4.7 \times 10^{-4}$ at $z\sim 0.3-0.4$,  the value which we use here.

{ Each of the contributions to the power spectrum -- viz. the one-halo, two-halo terms and a possible shot noise described by \eq{shotnoiseHI} -- are separately plotted for $z \sim 0.32$ in Fig. \ref{fig:hicontribs} as the solid lines. Overplotted are the MeerKAT intensity mapping observations \citep{paul2023} covering the $k$-range 0.4 - 8 $h$ Mpc$^{-1}$. 
}

{ Plotted in Figs. \ref{fig:psrsdmeerkat} are the full intensity mapping power spectra covering two redshift bins centred at $z = 0.32$ and $z = 0.44$ respectively (with the HI mass function at $z \sim 0.32$ presented in the previous section having been derived from the corresponding power spectrum). 
The dashed green lines in \cref{fig:psrsdmeerkat} show the power spectrum including only the 1- and 2-halo terms, in which only the 1-halo term contributes to the small scales. 

In the previous section, we have described the spherically averaged power spectra, implicitly assuming that the entire range of $\mu$ is available at each $k$ to sum or integrate over in \eq{rispherical}. This is not the case in the real measurement, with the foreground wedge leading to a loss of modes in the $k$-space power spectrum.
The foreground wedge is defined through the region in $k$-space enclosed by:
\begin{equation}
k_{\parallel, \rm min} = \frac{A}{\sqrt{1 + A^2}} k
\label{kpardef}
\end{equation}
where $A = 0.3$ is the foreground wedge factor adopted by the measurement. 
The maximum value $k_{\parallel, \rm max}$ is formally determined by the channel resolution of the survey. We have $\mu = k_{\parallel}/ \sqrt{k_{\parallel}^2 + k_{\perp}^2}$; and noting that $k_{\parallel}$ runs from 0.1 to $\sim 3$ Mpc$^{-1}$ and $k_{\perp}$ from 0.35 to $\sim 200$ Mpc$^{-1}$ \citep{paul2021} for MeerKAT, we obtain $\mu_{\rm min} = 0.014, \mu_{\rm max} = 0.27$.

In the presence of the foreground wedge as above, it can be shown that the expressions for $\mathcal{R}_i$ in \eq{rispherical} get modified to\footnote{ We note, however, that such a treatment is approximate since we do not additionally incorporate the weighting scheme employed in \citet{paul2023} in the conversion of the 3D power to its 1D version (and with which the results here may be degenerate).}:
\begin{align}
   {\cal R}_{1, \rm fw}(y) &= \frac{1}{\mu_{\rm max}-\mu_{\rm min}}\int_{\mu_{\rm min}}^{\mu_{\rm max}} d\mu \left(1+f\mu^2\right)^2 {\rm e}^{-{y}^2\mu^2}   
    \nonumber \\
   & = \frac{1}{\mu_{\rm max}-\mu_{\rm min}} \Big[\frac{\sqrt{\pi}}{8}\,\frac{ {\rm erf}(y \mu_{\rm max})}{y^5}\,
    \Big[ 3f^2+
    4fy^2+4y^4 \Big]  
     \nonumber\\
   &-  \frac{\mu_{\rm max}{\rm e}^{-y^2 \mu_{\rm max}^2}}{4y^4}\Big[ f^2\big(3+2y^2 \mu_{\rm max}^2\big)+4fy^2 \Big] 
    \nonumber\\
    &-\frac{\sqrt{\pi}}{8}\,\frac{ {\rm erf}(y \mu_{\rm min})}{y^5}\,
    \Big[ 3f^2+ 4fy^2+4y^4 \Big]  
    \nonumber \\
   &~~~~~
    + \frac{\mu_{\rm min}{\rm e}^{-y^2 \mu_{\rm min}^2}}{4y^4}\Big[ f^2\big(3+2y^2\mu_{\rm min}^2\big)+4fy^2 \Big]\Big]\\
   {\cal R}_{2, \rm fw}(y) =& \frac{1}{\mu_{\rm max}-\mu_{\rm min}}\int_{\mu_{\rm min}}^{\mu_{\rm max}} d\mu \,{\rm e}^{-{y}^2\mu^2/2}   %
   \nonumber \\
   &
   = \frac{1}{\mu_{\rm max}-\mu_{\rm min}}\sqrt{\frac\pi2}\,\frac{ {\rm erf}(y \mu_{\rm max}/\sqrt2\,) - {\rm erf}(y \mu_{\rm min}/\sqrt2\,)}y.
\end{align}
in which
$ \ y \equiv k\sigma$.
Using the above (wedge-corrected) versions of the $\sigma$-dependent terms in the power spectrum of Eqs. (\ref{rs}) - (\ref{p0}) leads to their foreground-corrected forms:
\begin{align}
\bar{P}_{\rm HI, fw}^{\rm s}(k) &= \left( F_{\rm HI, fw}^2+\frac23 F_{\rm m, fw}F_{\rm HI,fw} + \frac15 F_{\rm m, fw}^2\right) P_{\rm lin}(k) \notag\\
&+ \frac1{\bar{\rho}_{\rm HI}^2}
\int \d M \, n(M)  M_{\rm HI}^2 \,
  \big|u_{\rm HI}(k,M)\big|^2,
\label{rsfw}
\end{align}
where 
\begin{align}
F_{\rm m, fw}&=f\int \d M \, n(M)  b(M) \, {\cal R}_{2, \rm fw}(k\sigma)  u_{\rm h}(k, M), \nonumber\\
F_{\rm HI, fw}&=
\frac1{\bar{\rho}_{\rm HI}}  \int \d M \,
n(M)  M_{\rm HI}(M)  {b_{\rm h}}(M)\,  {\cal R}_{2,\rm fw}(k\sigma)  u_{\rm HI} (k,M)\;.
\label{p0fw}
\end{align}
plotted in the blue curves of Fig. \ref{fig:psrsdmeerkat} at $z \sim 0.32$ (top panel) and  $z \sim 0.44$ (lower panel).
 Noting that the one-halo term describes halo scales and thus line-of-sight effects and RSDs are not expected to be involved in it, we do not modulate the one-halo term by finger-of-god effects. 
We do not add any additional shot noise, interpreting it to be already present as the $k \to 0$ limit of the one halo term.}

The black data points with error bars show the MeerKAT measurements. { They are in line with the halo model predictions at $z = 0.32$, modified as above.  This is also borne out by the fact that a best-fit to the data, leaving free a scaling to the amplitude of the power spectrum, does not lead to significantly different results. Specifically, we apply such a correction only to the correct the temperature portion of the power spectrum derived from the HI density:
\begin{equation}
P_{\rm corr} (k) =  A_b^2 \bar{T}_{\rm HI}^2 \bar{P}_{\rm HI, fw}^{\rm s}(k) 
\label{pcorrfid}
\end{equation}
We find that the best-fitting value of $A_b = 1.13 \pm 0.13$, as enclosed by the solid black curve, with the uncertainty in $A_b$ covered by the grey shaded region. This shows that the amplitude hike is consistent with unity.}

At { the higher redshift interval $z = 0.44$}, the data favour a greater amplitude of the power  than predicted by the theory alone, which could have different possible causes. The above procedure, { fitted for $A_b$ from \eq{pcorrfid}} results in a best-fitting value $A_b = 1.55 \pm 0.08$, as shown by the black solid line and grey shaded region in { the lower panel of Fig. \ref{fig:psrsdmeerkat}} (note that the grey shading only shows the uncertainty in the fitting of the $A_b$ parameter).  This enhancement is in line with the findings of \citet{paul2023}, who impose a prior of $\Omega_{\rm HI} = 0.77 \times 10^{-3}$ for $z \sim 0.44$, the same factor higher than the value we assume here. } It also captures a possible remaining correction to the power which may result from higher-order biases not related only to the DM power as well as yet-unknown baryonic or other factors. The overall uncertainty in the fitting could be represented by the combined effect of the blue and grey shaded regions. 

\section{Discussion}
\label{sec:discussion}

We have examined the effects that modify the power spectrum of HI on small scales, using the latest results from the MeerKAT intensity mapping experiment to constrain the free parameters. In so doing, we have developed analytical fitting forms for the small scale structure of the HI intensity power spectra arising from the halo model framework. We have additionally included the contribution { of the foreground wedge caused by the non-spherical region in $k$-space in the  measurement.}

The behaviour of the HI mass - halo mass relation is consistent with the expectations of the halo model framework and also matches the findings of recent HI galaxy surveys in the relevant mass range \citep{bera2022}.
The measured power spectrum at $z=0.32$ is consistent with the expectations from the halo model, as also borne out by fitting the data directly leaving free an overall scaling to the amplitude of the power spectrum. Such a procedure at $z=0.44$ may indicate some evidence for a higher than expected amplitude corresponding to a constant increase to the HI bias that we obtained in the halo-model framework. This could have different possible causes, including higher-order biases not related to dark matter as suggested in SPT/EFT models. Beyond-EFT effects, such as the fact that the velocities of HI and dark matter do not agree on small scales and various complex baryonic effects may also contribute. { This finding is also indicative of a higher-than expected HI abundance at the latter redshift, in line with the prior adopted by \citet{paul2023} based on the DLA data at higher redshifts. This may indicate a more rapid evolution of the HI abundance than expected [though still within observational uncertainties, e.g., Fig. 2 of \citet{hptrcar2015}.]}

An additional consideration is the presence of nonzero vorticity in the dark matter, which adds a correction to the HI power spectrum. This correction is found via N-body simulations and can be modelled as  \citep{cusin2017}:
\begin{equation}
    P_{{\rm m},\omega} (k, z = 0)  =  \frac{k^2}{2\pi^3} \frac{A_\omega\,(k/k_*)^{n_l}}{\big[1 + (k/k_*)\big]^{n_l + n_s}} \, (\rm Mpc/h)^3\;.
\end{equation}
Typical values of the parameters $k_*, n_l$ and $n_s$ produce a peak in vorticity power at $k\sim1\,h/$Mpc.
The amplitude parameter $A_\omega$ is more uncertain. Most simulations find $A_\omega \sim 10^{-5}$, with some advocating a value 50-100 times larger \citep{bonvin2018, pueblas2009}. These values are found to be too low to explain the observed difference in the predicted and measured HI power in the current data.  { An improved fit also arises if the finger-of-god suppression $\sigma_{\rm avg}$ is artificially reduced to 1.0-1.4 Mpc, although this is inconsistent with the halo model framework -- which may also indicate the possibility of incompleteness of the framework.} A more detailed treatment of the these effects is left to future work (Umeh et al., in preparation).

At $z \sim 0.3$, the parameter $\Omega_{\rm HI}$ has been measured to be $4.8 \times 10^{-4}$ also from surveys of galaxies \citep{bera2019}, which is consistent with the value we use here, though as also suggested by the intensity mapping results, there may be evidence for an evolution with respect to mass \citep{bera2022}. At $z \sim 0.4$, the HI mass has been connected to the stellar mass of MeerKAT MIGHTEE galaxies, finding some evolution in the relationship out to $z \sim 1$. 

Future data sets would be useful to consolidate these findings to develop an deeper understanding of the role of HI in the baryon cycle for $z \sim 0$--$1$. The above formalism holds for any tracer of the dark matter which is in a mass-dependent form, so it would be equally applicable to small-scale effects in surveys probing CO/[CII]/[OIII] lines in the microwave and submillimetre regimes  \citep[GHz and THz frequencies, for reviews see, e.g., ][]{bernal2022, kovetz2019}. Thus far, the single-dish surveys used for measuring intensity fluctuations of these tracers lead to a suppression of the power spectrum on scales smaller than the beam size \citep[e.g.][]{hpoiii},  which is where most of the effects considered here become relevant.  This may,  however,  become possible with the availability of surveys probing smaller scales in other tracers in the future.

\section*{Acknowledgements}
\addcontentsline{toc}{section}{Acknowledgements}
We thank Sourabh Paul, Zhaoting Chen and Mario Santos for helpful discussions at an early stage of this work, { and the referee for detailed comments that improved the content and presentation}.
HP acknowledges support from the Swiss National Science Foundation via Ambizione Grant PZ00P2\_179934.
RM is supported by the South African Radio Astronomy Observatory and the National Research Foundation (Grant No. 75415). SC acknowledges support from the Italian Ministry of University and Research (\textsc{mur}) through PRIN 2022 `EXSKALIBUR – Euclid-Cross-SKA: Likelihood Inference Building for Universe's Research'.
OU  was partly supported by the 
UK Science \& Technology Facilities Council  Consolidated Grants Grant ST/S000550/1.

\section*{Data availability}
No new data were generated or analysed in support of this research. The software underlying this article will be shared on reasonable request to the corresponding author.

\bibliographystyle{mnras}
\bibliography{mybib} 

\appendix
\section{Contribution of satellite galaxies to the small-scale power}
\label{sec:satps}

\begin{figure}
    \centering
    \includegraphics[width = \columnwidth]{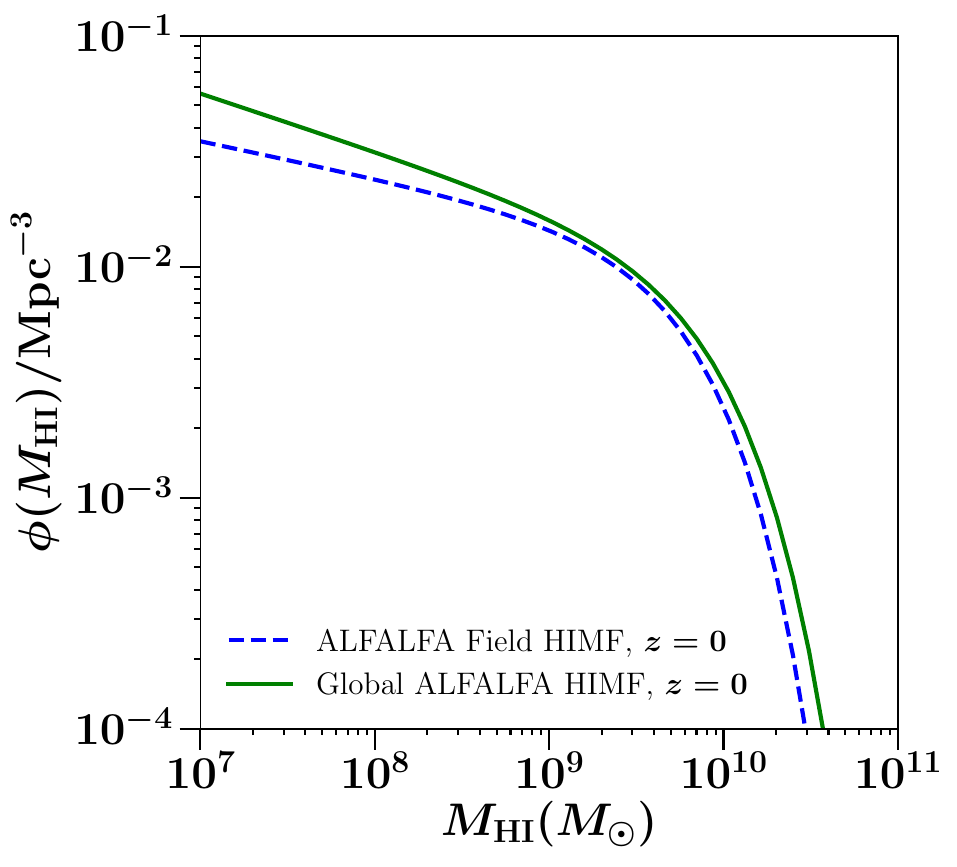} \includegraphics[width = \columnwidth]{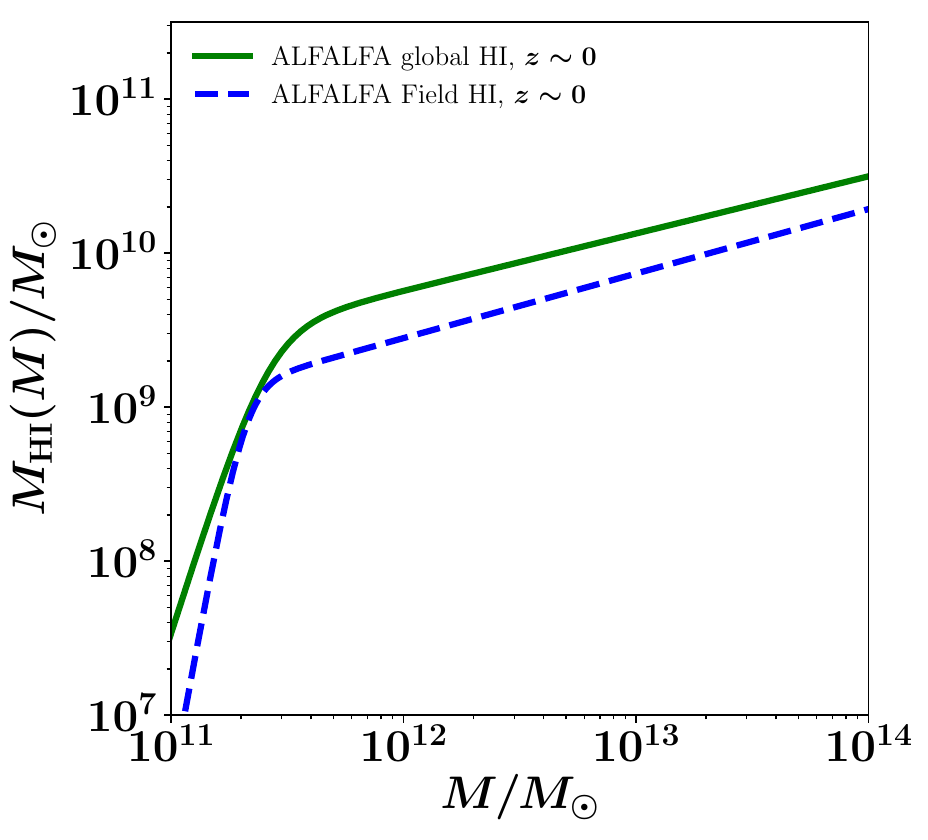}
    \caption{{{\it Top panel:} The HI mass function (HIMF) of Field galaxies (blue dashed lines) in the $z \sim 0$ ALFALFA survey, compared to the global HIMF (green solid line), which includes the group galaxies as well. {\it Lower panel:} The HI-to-halo mass relation inferred via abundance matching to the ALFALFA field galaxy HIMF, and that from the global HIMF.}}
    \label{fig:hialfalfa}
\end{figure}

Here, we provide an estimate of the contribution of the so-called `satellite' galaxies to the HI intensity mapping power spectrum on small scales.  As discussed in the main text, our analysis does not separate centrals and satellites explicitly, instead working with a continuous distribution of HI in haloes as is suitable in the context of intensity mapping. In the literature, there have been simulations on modelling HI in dark matter haloes with the explicit separation of central and satellite galaxies \citep[e.g.,][]{zhang2022, villaescusa2018, baugh2019}, with most analyses finding the satellite contribution to be negligible up to  halo masses of about $10^{12} M_{\odot}$. 

To investigate whether the satellite galaxies may be expected to make an impact on the small-scale structure of the HI power spectrum, we use the recent analyses of data from the Arecibo Legacy Fast ALFA survey, which provide the `global' HI mass function (HIMF) of galaxies in the full sample \citep{jones2018}. By removing the HI detections within known galaxy groups of the Sloan Digital Sky Survey (SDSS), the survey also provides the HIMF of the remaining galaxies, labelled the `Field HIMF' \citep{jones2020}. Both HIMFs are found to be well-described by Schechter functions, of the form:
\begin{equation}
\phi(M_{\rm HI}) = \ln(10) \phi_* \left(\frac{M_{\rm HI}}{M_*}\right)^{\alpha + 1} \exp -\left(\frac{M_{\rm HI}}{M_*}\right)
\end{equation}
with the parameters $\alpha = -1.25, \log_{10} (M_*/M_{\odot}) = 9.94$, and $\phi_* = 4.5 \times 10^{-3} $ Mpc$^{-3}$ dex$^{-1}$ (for the global HIMF) and $\alpha = -1.16, \log_{10} (M_*/M_{\odot}) = 9.81 $, and $\phi_* = 5.4 \times 10^{-3} $ Mpc$^{-3}$ dex$^{-1}$ (for the Field HIMF).
Both mass functions are plotted in the top panel of Fig. \ref{fig:hialfalfa}.

As previously \citep[e.g.,][]{hpgk2017}, we use the abundance matching technique to infer the HI to halo mass (HIHM) relation directly from both HIMFs. The difference between the HIHMs quantifies the contribution of group galaxies as a function of halo mass. Both HIHMs are found to be describable by double power laws:
\begin{equation}
M_{\rm HI} (M) = 2N M [(M/M_1)^{-b} + (M/M_1)^{y}]^{-1}
\end{equation}
as shown in the lower panel of Fig. \ref{fig:hialfalfa}. As expected, the contribution of group galaxies dominates at high halo masses.
The best fitting parameters of the HIHMs are found to be $M_1 = 2.65 \times 10^{11} M_{\odot} , N = 0.0066 , b = 3.73$, and $y = 0.63$ (global) and $M_1 = 2.09 \times 10^{11} M_{\odot} , N = 0.0035 , b = 7.20$, and $y = 0.58$ (Field).
The similarity of the HIHM relations re-illustrates that the vast majority of HI exists in central galaxies, with a negligible contribution of satellites especially at low masses, as found by \citet{jones2020}.

The shot noise term calculated by using \eq{shotnoiseHI} comes out to be 256 Mpc$^{3}$ for the global HIMF and 260 Mpc$^{3}$ for that of the field galaxies, confirming the negligible contribution of satellites at the small scales.

It is also possible to calculate the contribution of group galaxies to the HIMF by using the `corrected' group HIMF introduced in \citet{jones2020}, in which the group and field HIMFs are multiplied by the volume they correspond to, giving the total corrected number counts per dex. The corrected counts can be directly compared in view of the difference in the physical volumes occupied by the group and field galaxies. It is seen (cf. Fig. 8 of \citet{jones2020}) that the above correction to the field HIMF is between a factor of 1.13 and 7.09 as the HI mass is varied from 10$^7$ to 10$^{11}$ $M_{\odot}$. The corrected HIMF can be used to infer an HIHM as above using the abundance matching technique, which results in a shot noise of 252 Mpc$^{3}$ for the `group-corrected' field catalogue. This re-iterates the conclusion that the contribution of satellite galaxies to the HI power spectrum is negligible at small scales. At the redshifts under consideration in the present paper, this contribution is expected to decrease further, since large halo masses are down-weighted by HI contribution with increase in redshift.

\section{The two-dimensional power spectrum}
\label{sec:2dps}

\begin{figure}
    \centering
    \includegraphics[width = 0.97 \columnwidth]{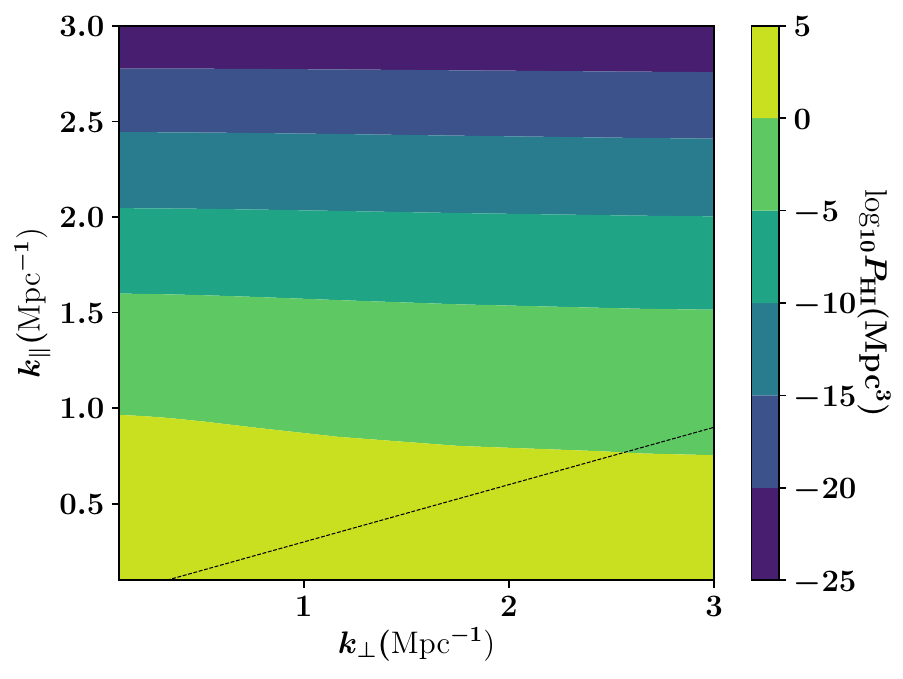} \includegraphics[width = 0.97 \columnwidth]{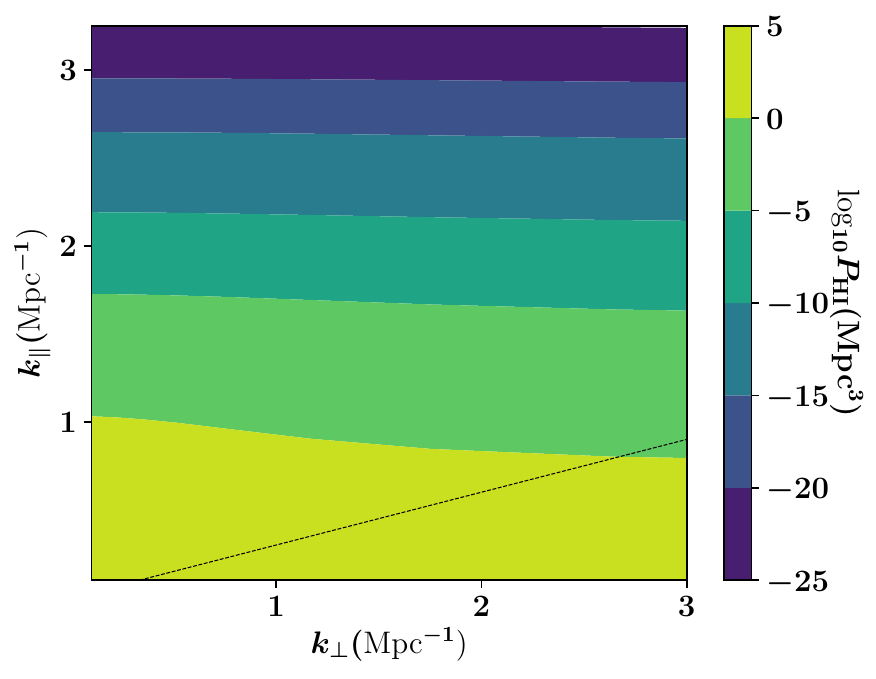}
    \caption{Two dimensional power spectra in redshift space with the parameters  following the halo model prescription of \citet{hparaa2017} and the Gaussian finger-of-god effects as discussed in the main text, at $z =0.32$ (top) and $z = 0.44$ (lower panel). { The horizon limit adopted by the MeerKAT results is shown in each figure by the black dotted line}.
    }
    \label{fig:2dpshp}
\end{figure}

The two-dimensional power spectrum \citep[e.g.,][]{paul2023}, i.e.  in the $(k_{\perp},k_{\parallel})$ plane,  
{contains more information than the} 
monopole of the power spectrum considered in \cref{sec:hismallscales}:
\begin{equation}
P_{\rm HI}^{\rm s}(k_{\parallel}, k_{\perp}) = P_{\rm HI} ({k_{\parallel}, k_{\perp}}) \bigg(1 +  \frac{k_{\parallel}^2}{k_{\parallel}^2+k_{\perp}^2} \, \frac{f}{b_{\rm HI}}\bigg)^2 
%D_{\rm fog}(\roy{k_{\parallel}}).
\exp\big({-{k_{\parallel}^2} \sigma_{\rm avg}^2}\big).
%~~\beta=\frac{f}{b_{\rm HI}}.
\label{2dpowerHI}
\end{equation}
{ where $\sigma_{\rm avg}$ is given by \eq{savg}.}
The resultant 2D power spectra are displayed in \cref{fig:2dpshp}, for the two redshifts, $z=0.32,\, 0.44$. { These are not affected by the window function or the $k$-space asymmetry in real observations and are thus directly comparable with future data}.  The plots show that the power rapidly declines (by several orders of magnitude) as $k_{\parallel}$ increases, due to the finger-of-god effects at high-$k$. This follows from the value of $\sigma_{\rm avg} \sim 3.5$ Mpc at these redshifts, combined with the exponential fall-off in \eq{2dpowerHI}.

The wedge limit of the MeerKAT interferometer, below which foregrounds are expected to become relevant, is defined through 
\citep[e.g.][]{liu2014}:
\begin{equation}
k_{\parallel} = \frac{r(z) {H}(z) \sin \theta_0}{c (1+z)}\, k_{\perp} {\simeq 0.02\, k_{\perp}}\,.
\end{equation}
Here $r$ is the comoving radial distance and {$\theta_0$} is a characteristic of the beam  \citep{paul2023}.
The wedge adopted in the  analysis of \citet{paul2023} is taken to be about 10 times stricter,  $k_{\parallel} = 0.3 k_{\perp}$. This is shown by the black dotted lines in Fig. \ref{fig:2dpshp}.

\bsp	
\label{lastpage}
\end{document}